\documentclass{aa}

\usepackage{graphicx}
\usepackage{txfonts}
\usepackage{xcolor}
\usepackage{xspace}
\usepackage{hyperref}
\usepackage{comment}
\usepackage{multirow}
\usepackage{amsmath}
\usepackage{mathrsfs}
\hypersetup{
     colorlinks   = true,
     citecolor    = blue
}

\def \integral {{\em INTEGRAL}}

\def \xmm {{\em XMM-Newton}}
\def \rxte {{\em RXTE}}
\def \nicer {{\em NICER}}

\def \igr{{\em IGR J17379$-$3747}}
\def\swiftj{{\em Swift J1756.9$-$2508}}
\def\saxJ{{SAX J1748.9$-$2021}}
\def\saxj{SAX J1808.4$-$3658}

\def \Msun{{M$_{\odot}$}}

\begin{document}

\title{\xmm{} detection of the 2.1~ms coherent pulsations from \igr{}}

   \author{A. Sanna\inst{1},
          E. Bozzo\inst{2},
          A. Papitto\inst{3},
          A. Riggio\inst{1},
          C. Ferrigno\inst{2},
          T. Di Salvo\inst{4},
          R. Iaria\inst{4},\\
          S. M. Mazzola\inst{4},
          N. D'Amico\inst{1,5},
          L. Burderi\inst{1}.}

   \institute{Dipartimento di Fisica, Universit\`a degli Studi di Cagliari, SP Monserrato-Sestu km 0.7, 09042 Monserrato, Italy\\
   		\email{andrea.sanna@dsf.unica.it}
   	\and
    	ISDC, Department of Astronomy, University of Geneva, Chemin d'\'Ecogia 16, CH-1290 Versoix, Switzerland
         \and
         INAF, Osservatorio Astronomico di Roma, Via di Frascati 33, I-00044, Monteporzio Catone (Roma), Italy
         \and
               Universit\`a degli Studi di Palermo, Dipartimento di Fisica e Chimica, via Archirafi 36, 90123 Palermo, Italy  
         \and
                 INAF, Osservatorio Astronomico di Cagliari, Via della Scienza 5, I-09047 Selargius (CA), Italy        
             }

   \date{Received -; accepted 19 July 2018}

  \abstract
   {We report on the detection of X-ray pulsations at 2.1 ms from the known X-ray burster \igr{} using \xmm{}. The coherent signal shows a clear Doppler modulation from which we estimate an orbital period of $\sim$1.9 hours and a projected semi-major axis of $\sim8$ lt-ms. Taking into account the lack of eclipses (inclination angle of $<75^{\circ}$) and assuming a neutron star mass of 1.4~\Msun, we estimated a minimum companion star of $\sim 0.06$~\Msun. Considerations on the probability distribution of the binary inclination angle make less likely the hypothesis of a main-sequence companion star. On the other hand, the close correspondence with the orbital parameters of the accreting millisecond pulsar SAX J1808.4-3658 suggests the presence of  a bloated brown dwarf. The energy spectrum of the source is well described by a soft disk black-body component (kT $\sim$0.45 keV) plus a Comptonisation spectrum with photon index $\sim$1.9. No sign of emission lines or reflection components is significantly detected. Finally, combining the source ephemerides estimated from the observed outbursts, we obtained a first constraint on the long-term orbital evolution of the order of $\dot{P}_{\rm orb}=(-2.5\pm2.3)\times 10^{-12}$s/s.}

  \keywords{X-rays: binaries; stars:neutron; accretion, accretion disc, \igr{}
               }

\titlerunning{2.1~ms X-ray pulsations from \igr{}}
\authorrunning{Sanna et al.}

   \maketitle

\section{Introduction}

Accreting millisecond X-ray pulsars (AMXPs) are quickly rotating neutron stars (NS) which accrete mass transferred from a low mass ($\leq M_{\odot}$) companion star via Roche lobe overflow. The observations of X-ray pulsations at the NS spin period shows that the magnetic field of the NS in these systems is strong enough to channel the mass flow to the magnetic poles. The discovery of AMXPs \citep{Wijnands1998a} demonstrated that a prolonged phase of mass accretion is able to spin up a NS to such a quick rotation. When mass transfer ceases a rotation-powered radio millisecond pulsar (msp) turns on \citep{Alpar82}. The discovery of transitional msp that are able to switch between accretion and rotation-powered regimes \citep{Archibald2009a,Papitto2013b} has recently demonstrated the tight evolutionary link shared by neutron stars in low mass X-ray binaries and radio msp. 
\noindent
Twenty AMXPs have been discovered, so far \citep[e.g.,][]{Burderi13, Patruno2017, Sanna2018}, a small fraction of the $>200$ low mass X-ray binaries known to host a NS. They are all relatively faint X-ray transients which attain a luminosity of 0.01-0.1 Eddington rate at the peak of their outbursts, which are typically a few weeks-long. X-ray pulsations at a period ranging between 1.6~ms \citep{Galloway05a} to 9.5~ms \citep{Sanna2018} are observed during the X-ray outburst with a fractional amplitude of the order of $\sim1-10\%$. AMXPs have been found in binary systems with an orbital period shorter than a day and as low as $\sim40$~minutes, indicating companion stars with masses ranging from $~0.6$ to $0.01\,M_{\odot}$ \citep[see e.g.][]{Patruno12b, Burderi13}.

\igr{} is a known X-ray burster observed in outburst by the {\rm Rossi X-ray Timing Explorer} (\rxte{}) in February 2004 and September 2008 \citep{Markwardt2008,Shaw2008,Krimm2008} and designated XTE J1737-376. The source position was consistent with a faint source reported in the third IBIS/ISGRI soft gamma-ray survey catalogue \citep{Bird:2007aa}, \igr{}. The peak flux attained during the outbursts ranged between $1.2$ and $2.5\times10^{-10}$~erg~cm$^{-2}$~s$^{-1}$ (2-10 keV), corresponding to a luminosity of $1-2\times10^{36}$~erg~s$^{-1}$ assuming a distance d=8.5~kpc. The spectral energy distribution was described by a power law with $\Gamma\simeq1.8$--$2.2$ \citep[see e.g.][]{Strohmayer2018,Eijnden2018,Eijnden2018b}. Renewed activity from the source was detected by MAXI/GSC on March 19, 2018 \citep{Negoro2018} at a 4-10~keV flux of $(4\pm1)\times10^{-10}$~erg~cm$^{-2}$~s$^{-1}$. Very Large Array (VLA) observations performed during the 2018 outburst by \citet{Eijnden2018} detected a flat-spectrum radio counterpart with a flux density of $\simeq0.4~mJy$ at 4.5 and 7.5~GHz at a position consistent with the X-ray source. 
X-ray coherent pulsation at $\sim468$ Hz has been reported from observations performed by \nicer{} \citep{Strohmayer2018}. In this letter, we report on the \xmm{} observation performed on 2018 April 1, almost a week after the beginning of the outburst.

\section[]{Observations and data reduction}

\subsection{\xmm{}}
\xmm{} \citep{Jansen2001} performed a dedicated target of opportunity observation of \igr{} (Obs.ID. 0830190301) on 2018 April 1 at 02:12 UTC for an elapsed time of $\sim45$ ks. Few ks after the beginning of the observation a type-I burst was detected with a peak flux of $\sim1.6\times$10$^{-8}$~erg~cm$^{-2}$~s$^{-1}$  (more details on this event will be published elsewhere).

During the observation the Epic-pn (PN) camera was operated in timing mode, Epic-MOS 1-2 in timing mode, while the RGS observed in spectroscopy mode. 
We filtered the PN and MOS data using the Science Analysis Software (SAS) v. 16.1.0 with the up-to-date calibration files, and following the standard reduction pipeline \textsc{epproc} and \textsc{emproc}, respectively. We filtered the source data in the energy range 0.3-10 keV, selecting events with \textsc{PATTERN$\leq$4} and \textsc{(FLAG=0)} to retain only events optimally calibrated for spectral analysis. We selected source and background events from the PN regions RAWX [27:47] and [2:8], respectively. The average source spectral properties were analysed by removing the type-I burst detected at the beginning of the observation and integrating the PN and MOS2 spectra during the remaining observational time. Following standard analysis procedure, we generated in all cases the response matrix and the ancillary file using the \textsc{rmfgen} and \textsc{arfgen} tools, respectively. Energy channels have been grouped by a factor of three to take into account the oversampled energy resolution of the instrument, and we also binned the energy spectrum to guarantee at least 25 counts per bin. We discarded both MOS1 (due to the ``hot column'' issue\footnote{\url{https://www.cosmos.esa.int/web/xmm-newton/sas-watchout-mos1-timing}}) and RGS data because of the poor signal-to-noise ratio (S/N).

For the timing analysis we reported the PN photon arrival times to the Solar System barycentre by using the \textsc{barycen} tool (DE-405 solar system ephemeris) adopting the best available source position obtained from the VLA observation of the source \citep{Eijnden2018}, and reported in Tab.~\ref{tab:solution}.

\subsection{\rxte{}}
\igr{} has been observed by \rxte{} during its enhanced X-ray activity phases in 2004 (Obs.ID. 80138-07-01-00, for a total exposure of $\sim 4$ ks) and in 2008 (Obs.IDs. 93044-14-01-00, 93449-01-02-00/01/02/03/04/05 and 93449-01-03-00, for a total exposure of 14 ks). To perform the timing analysis we selected data collected by the PCA \citep{Jahoda96} in Event packing mode with time resolution $\leq 500\mu$s, that we processed and analysed using the \textsc{HEASARC FTOOLS} v.6.23. To improve the S/N ratio, we selected photon events in the energy range 3-15 keV. We used the \textsc{faxbary} tool to apply barycentric corrections.

\subsection{\integral{}}

Following the detection of a new outburst from \igr{}, a 50 ks-long \integral{} observation was performed from 2018 April 1 at 08:30 to 23:15 (UTC). We analysed all available ``science windows'' with the OSA 10.2 software distributed by the ISDC \citep{courvoisier03}. We used  IBIS/ISGRI \citep{lebrun03,ubertini03} and JEM-X \citep{lund03} data. As the observation was carried out during the rapid decay of the source outburst, \igr{} was not significantly detected in the IBIS/ISGRI and JEM-X mosaics. We estimated 3~$\sigma$ upper limits on the source flux of 5$\times$10$^{-11}$~erg/cm$^2$/s in the 20-100 keV energy band and 1.0$\times$10$^{-10}$~erg/cm$^2$/s in the 3-20 keV energy band. These are compatible with the flux obtained from the \xmm{} observation started about 7 hours before the \integral{} pointings (Sect.~\ref{sec:spectral}). No evidence for type-I burst has been found in the JEM-X light curves.

\section{Data analysis}

\subsection{Timing analysis}
\label{sec:ta2017}

\begin{figure}
\centering
\includegraphics[width=0.4\textwidth]{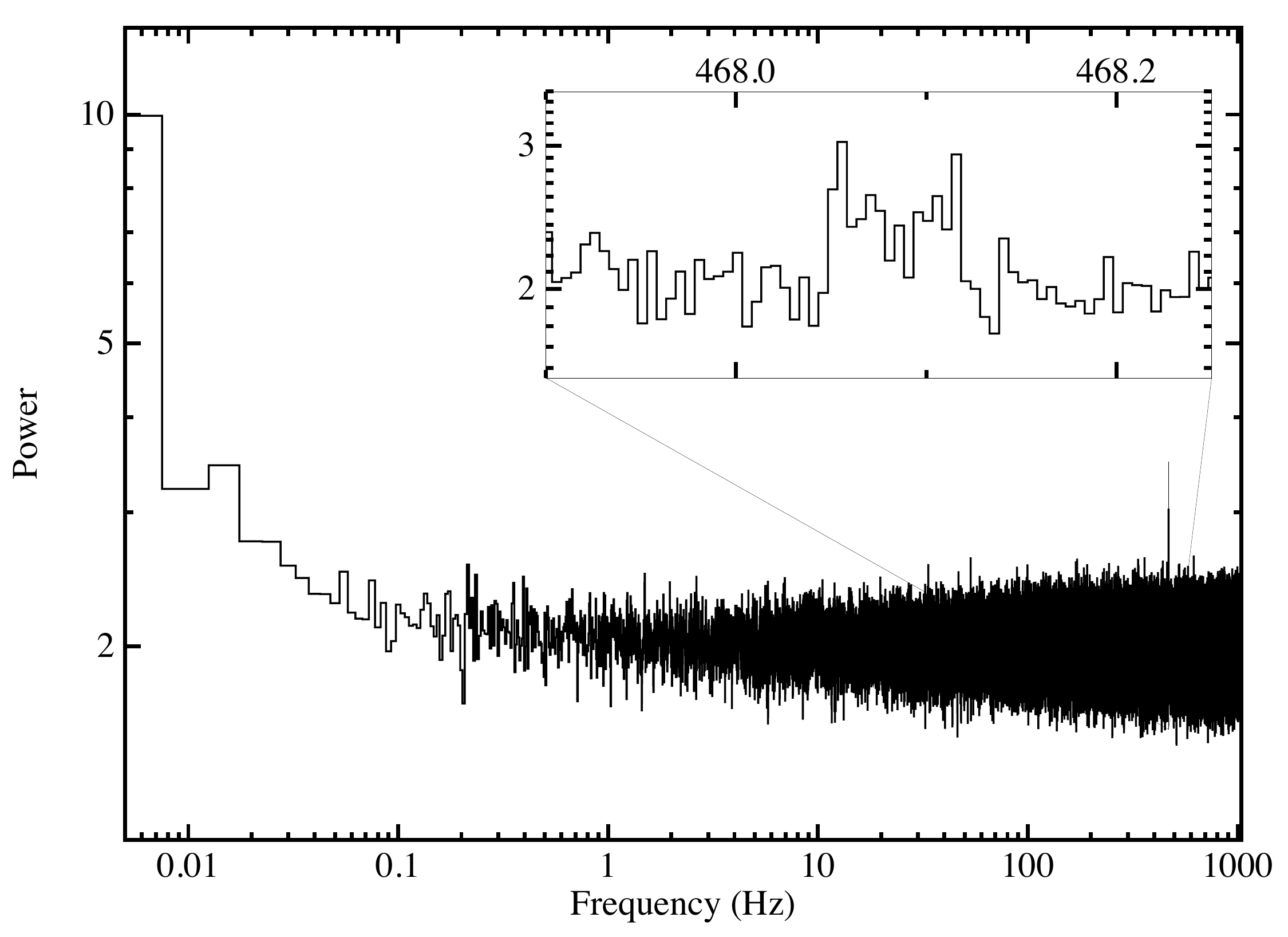}
\caption{Leahy normalised \citep{Leahy1983a} PDS, produced by averaging 200 s-long segments of \xmm{} data. A statically significant coherent signal is visible at a frequency of $\sim468$ Hz. The inset shows a zoom of the frequency double-peaked profile.}
\label{fig:pds}
\end{figure}

We search for coherent signals by generating a power density spectrum (PDS) averaging 225 power spectra produced on 200-s data segments from the \xmm{} observation  (Fig.~\ref{fig:pds}; the type-I burst was removed before performing the timing analysis).  A double-peaked feature associated with orbital Doppler shift is significantly detected (>5$\sigma$) with a central frequency of $\sim 468.08$ Hz and a width of $\sim6\times 10^{-2}$ Hz (see inset in Fig.~\ref{fig:pds}). To investigate binary properties, we created PDS every 1000-s and inspected them for coherent features in the $\sim 6\times 10^{-2}$ Hz interval around the mean peak frequency previously mentioned. Assuming binary Doppler shift, we modelled the time evolution of the spin frequency obtaining the preliminary orbital solution: 
$\nu_0=468.0801(19)$ Hz, $x = 0.064(6)$ lt-s, $P_{\rm orb}=6699(50)$ s,  $T_{\rm NOD}=58208.965(3)$ (MJD), where $T_{\rm NOD}$ is the time of passage through the ascending node, $P_{\rm orb}$ is the orbital period and $x$ is the projected semi-major axis of the NS orbit in light seconds. We note that this timing solution is consistent with the preliminary source ephemeris reported by \citet{Strohmayer2018}.

Starting from these orbital parameters we corrected the photon time of arrivals for the binary motion through the recursive formula valid in the approximation of nearly circular orbits (eccentricity $e \ll 1$) $[t + x\,\sin M = t_{\rm arr}]$, where $M=2\pi(t-T_{\rm NOD})/P_{\rm orb}$, $t$ is the photon emission time and $t_{\rm arr}$ is the photon arrival time to the Solar System barycentre. We then iterated the process previously described, searching for coherent signals in PDS created in 500-s data segments. We modelled the residual spin frequency variation of the signal by fitting differential correction to the orbital parameters. We repeated the process until no significant differential corrections were found for the parameters of the model. The most accurate set of orbital parameters obtained with this method are: $\nu_0=468.08328(2)$ Hz, $x = 0.07694(5)$ lt-s, $P_{\rm orb}=6765.9(4)$ s, $T_{\rm NOD}=58208.96642(2)$ (MJD).

To further investigate the spin frequency and the orbital parameters, we then epoch-folded segments of $\sim$500 seconds in 8 phase bins at the frequency $\nu_0=468.08328(2)$ Hz obtained earlier. We modelled each pulse profile with a sinusoid of unitary period, and we determined the corresponding amplitude and the fractional part of the epoch-folded phase residual. 
We filtered pulse profiles such that the ratio between their amplitude and the corresponding  1~$\sigma$ uncertainty were equal or larger than three. To model the temporal evolution of the pulse phase delays we define the function $\Delta \phi(t)=\phi_0+\Delta \nu_0\,(t-T_0)+R_{\rm orb}(t)$, where $T_0$ represents the reference epoch for the timing solution, $\Delta \nu_0$ is the  spin frequency correction and $R_{\rm orb}$ is the R{\o}mer delay \citep[e.g.][]{Deeter81}. The described process was iterated until no significant differential corrections were obtained. Best-fit parameters are shown in Tab.~\ref{tab:solution}. The best pulse profile obtained by epoch-folding the whole \xmm{} dataset is well described by the superposition of three sinusoids, where the fundamental, second and third harmonics have fractional amplitudes (background corrected) of 
$\sim13\%$, $\sim3\%$ and $\sim1.3\%$, respectively.

We searched for coherent X-ray pulsations in the data collected with \rxte{} during the 2004 and 2008 outbursts of the source. For each outburst, we searched for the best local $T_{\rm NOD}$ (keeping fixed $P_{\rm orb}$ and $x$ to the values reported in Tab.~\ref{tab:solution}), applying epoch-folding search techniques to the data for each trial $T_{\rm NOD}$. We detected pulsations in both the 2004 and 2008 outbursts with a statistical significance of 6$\sigma$ and 5$\sigma$ (single trial) at a barycentric frequency of 468.08332(25) Hz and 468.0831(2) Hz, and local $T_{\rm NOD}=53056.03926(12)$ and $T_{\rm NOD}=54721.03253(11)$ MJD, respectively. The pulse profiles are well described by a sinusoid with (background corrected) fractional amplitudes of $3.6(5)\%$ and $4.5(4)\%$, respectively.

Finally, combining the orbital ephemerides measured for the three outburst of the source, we investigated the orbital period secular evolution by studying the delay accumulated by $T_{\rm NOD}$ as a function the orbital cycles elapsed since its discovery. To make sure of the feasibility of the coherent (orbital) timing analysis we verified the condition 
\begin{equation}
\left(\sigma^2_{T_{\text{NOD}}}+\sigma^2_{P_{orb}} N^2+\frac{1}{4}P^2_{orb} \dot{P}^2_{orb} N^4\right)^{1/2} \ll \frac{P_{orb}}{2}, 
\label{eq:orb_cycles}
\end{equation}

where $\sigma_{T_{\text{NOD}}}$ and $\sigma_{P_{orb}}$ are the uncertainties on the time of passage from the ascending node and the orbital period used to create the timing solution, respectively. $\dot{P}_{orb}$ represents the secular orbital derivative while $N$ corresponds to the integer number of orbital cycles elapsed by the source during the time interval of interest. To verify Eq.~\ref{eq:orb_cycles} we considered the best available estimate of the orbital period $P_{orb}=6765.90(2)$s obtained from the timing analysis of the \nicer{} observations of the latest outburst of the source (Markwardt et al. 2018 in prep.). Moreover, to overcome the lack of knowledge on the orbital period derivative (see e.g. Sanna et al. 2018 submitted), we considered the absolute value of average on the only two estimates reported for AMXPs \citep[$\dot{P}_{orb}=3.6(4)\times 10^{-12}$  s/s for \saxj{} and $\dot{P}_{orb}=1.1(3)\times 10^{-10}$  s/s for \saxJ{}, see e.g.][]{diSalvo08, Patruno:2017aa, Sanna:2017ab,Sanna2016a}, corresponding to $\dot{P}_{orb}\leq |6|\times 10^{-11}$  s/s. With these values for the orbital period and orbital period derivative, we found that Eq.~\ref{eq:orb_cycles} is verified only between the first two outbursts of the source. Phase connecting the orbital solutions of the 2004 and 2008 outbursts of \igr{} we obtained an improved estimate of the orbital period ${P}_{orb,04-08}=6765.845(4)$s. Adopting the more accurate estimate of the orbital period and assuming the same prescription for $\dot{P}_{orb}$, we then obtained that Eq.~\ref{eq:orb_cycles} is verified for the 2004-2018 baseline that includes the three outbursts investigated. To determined the delay accumulated by $T_{\text{NOD}}$ we estimated the expected $T_{\text{NOD}}$ for constant orbital period, $T_{\text{NOD,PRE}}(N)=T_{\text{NOD,04}}+N\, P_{orb,04-08}$, where $T_{\text{NOD,04}}$ is the time of passage from the ascending node observed during the 2004 outburst. For each outburst we estimated the quantity $T_{\text{NOD,obs}}-T_{\text{NOD,PRE}}$, and we modelled its evolution as a function of the elapsed orbital periods with the quadratic function: 
\begin{eqnarray}
\label{eq:fit_tstar}
\Delta T_{\text{NOD}} = \delta T_{\text{NOD,04}} + N\, \delta P_{orb,04-08}+0.5\,N^2\, \dot{P}_{orb}P_{orb,04-08},
\end{eqnarray} 
where $\delta T_{\text{NOD,04}}$ represents the correction to the adopted time of passage from the ascending node and $\delta P_{orb,04-08}$ is the correction to the orbital period. We obtained an improved estimate of the orbital period $P_{orb}=6765.84521(3)$ s as well as a first constraint on the orbital period derivative $\dot{P}_{orb}=(-2.5\pm 2.3)\times 10^{-12}$ s/s, where the uncertainties are reported at the $1\sigma$ confidence level.

\begin{table}
\scriptsize
\centering
\caption{Spin frequency and orbital ephemeris of \igr{} estimated during the three observed outbursts. 
Errors are at 
1$\sigma$ confidence level.$^a$ This parameter has been fixed to the value obtained from the \xmm{} timing solution.} 
\begin{tabular}{l | c c c }
Parameters &  2018 (\xmm{}) & 2004 (\rxte{}) & 2008 (\rxte{})\\
\hline
\hline
R.A. (J2000) &  \multicolumn{3}{c}{$17^h37^m58.836^s \pm 0.002^s$}\\
DEC (J2000) & \multicolumn{3}{c}{$-37^\circ46\arcmin 18.35\arcsec \pm 0.02\arcsec$}\\
\hline
$P_{\rm orb}$ (s) &6765.6(1)& 6765.6$^a$ & 6765.6$^a$\\
$x$ (lt-s) &0.07699(1)& 0.07699$^a$ & 0.07699$^a$\\
$T_{\rm NOD}$ (MJD) & 58208.966437(9)& 53056.03926(12) & 54721.032403(35)\\
e &$<1\times 10^{-3}$& --& --\\
  $\nu_0$ (Hz) &468.0832666(3) & 468.08338(13) & 468.0831(2) \\
$T_0$ (MJD) & 58209.0 & 53056.0 & 54721.0 \\
\hline
$\chi^2_\nu$/d.o.f. & 60.4/61  & -- & -- \\
\end{tabular}
\label{tab:solution}
\end{table}


\subsection{Spectral analysis}
\label{sec:spectral}
\subsubsection{The average spectrum} 
\begin{figure}
  \includegraphics[angle=-90, width=0.5\textwidth]{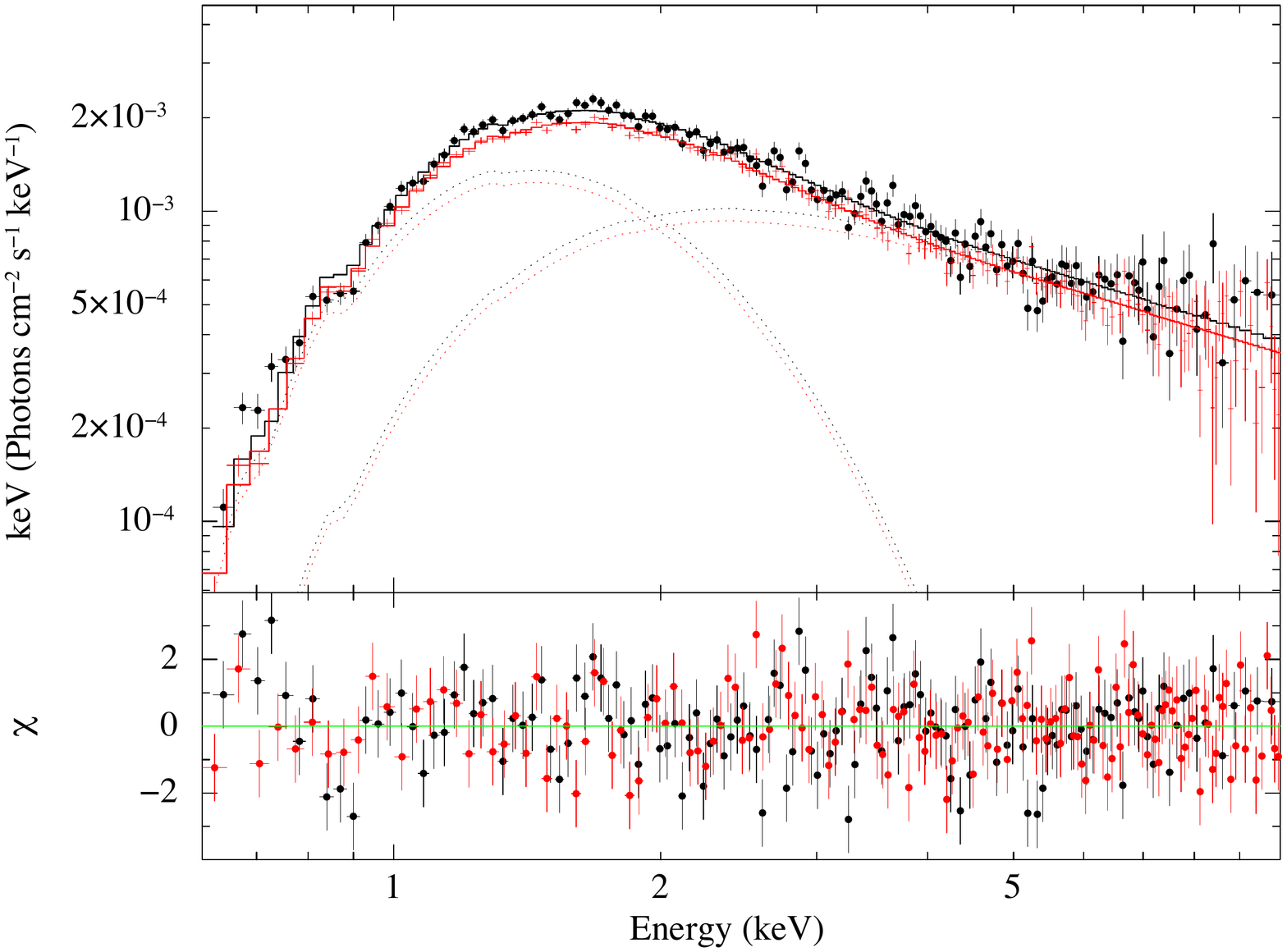}
  \caption{Unfolded Epic-pn (red) and MOS2 (black) average spectra of \igr{}. The residuals from the best fit described in the text are reported in the bottom panel.}      
  \label{fig:spectrum}
\end{figure}

We performed spectral analysis with Xspec 12.10.0 \citep{Arnaud96} and fit the average PN and MOS2 spectra in the 0.5-10 keV range (Fig.~\ref{fig:spectrum}). We assumed \citet{Wilms00} elemental abundances and \citet{Verner96} photo-electric cross-sections to model the interstellar medium. We allowed for a normalisation coefficient between instruments, to account for cross-instrument calibration offsets.

The spectra are well-described ($\chi^2_{\rm red}$/d.o.f.=1.13/282) by the typical model used for AMXPs in outburst, comprising an absorbed disk black-body plus a thermally comptonised continuum with seed photons from the black-body radiation ({\texttt const$\times$tbabs$\times$[diskbb+nthcomp]} in Xspec). We measured an absorption column density of $(0.90\pm0.03)\times10^{22}$~cm$^{-2}$, consistent with that expected in the direction of the source\footnote{\url{https://heasarc.gsfc.nasa.gov/cgi-bin/Tools/w3nh/w3nh.pl?}}. We obtained from the fit an inner disk temperature of $0.45\pm0.03$ keV (linked to be the same of the seed photon temperature in the Comptonisation component), and a photon index of $1.88\pm0.08$. The inter-calibration constant of the MOS2 with respect to the PN was found to be 1.08$\pm$0.01. The electron temperature of the {\texttt nthcomp} component could not be constrained in the fit and we fixed it to 30 keV, following the results obtained from similar sources \citep{Sanna2018}. We verified that using reasonably different values of this parameter did not affect the overall fit results. No evidence for spectral lines (e.g., Iron K-$\alpha$) was found by inspecting the residuals from the best fit to the PN and MOS2 data. Considering the iron line properties observed in AMXPs \citep[see e.g.][Pintore et al. 2018 submitted]{Papitto09,Papitto2013a,Papitto:2016aa, Pintore2016a, Sanna2017a} such as line energy in the range 6.4--6.97 keV and line width between 0.1--0.7 keV, we estimated an upper limit on the equivalent width of any iron line not detected during the source outburst ranging between 50--400 eV, still compatible with the lines observed in other AMXPs. The average 0.5-10~keV flux measured from the spectral fit was of (1.16$\pm$0.02)$\times$10$^{-11}$~erg~cm$^{-2}$~s$^{-1}$. 

\section{Discussion}
We report on the detection of X-ray coherent pulsations at $\sim 468$ Hz from the known type-I burster \igr{} observed by \xmm{} during its 2018 outburst. Coherent timing analysis of the ms pulsation allowed to determine properties of the binary system such as its $\sim 1.9$h orbital period and the $\sim0.08$lt-s NS projected semi-major axis. The orbital ephemeris reported here are consistent with those independently determined with the data collected by the \nicer{}'s X-ray Timing Instrument between 2018 March 29 and April 1 \citep{Strohmayer2018}.  

Combining the binary system mass function $f(m_2, m_1, i)\sim8\times 10^{-5}$~M$_{\odot}$ with the absence of eclipses (inclination angle of $i \lesssim 75^{\circ}$) in the 
X-ray light curve, we constrained the mass of the donor star to be $m_2 \gtrsim 0.056$~M$_{\odot}$ assuming a NS mass of 1.4~M$_{\odot}$ ($m_2 \gtrsim 0.07$~M$_{\odot}$ for a 2~M$_{\odot}$ NS). Assuming a Roche-Lobe filling donor star and fixing the NS mass, we can compare the companion mass-radius relation with that of theoretical H main-sequence stars \citep[e.g.][]{Tout1996a}. We find that, for a 1.4~M$_{\odot}$ NS, main sequence stars with $m_2 > 0.16$~M$_{\odot}$ would have a radius equal or larger than the donor Roche-Lobe, implying an inclination angle $i\leq21^{\circ}$. For an isotropic a priori distribution of the binary inclination angles, the probability to observe a system with $i\leq21^{\circ}$ (hence $m_2 > 0.16$~M$_{\odot}$) is $\leq 7\%$. In analogy with other AMXPs with very similar orbital parameters, such as SAX J1808.4-3658 and IGR J00291+5934, the companion star could be a hot brown dwarf, likely heated by low-level X-ray radiation during the quiescence phases \citep[e.g.,][]{Bildsten2001a, Galloway05a}.   

The average energy spectrum of \igr{} analysed here is well described by a superposition of a soft disk component ($kT\sim 0.45$ keV) and a hard power-law ($\Gamma \sim 1.9)$, consistent with typical AMXP observed in outburst \citep[e.g.][]{Gierlinski2005a, Papitto09,Falanga2013a}. We found no evidence for emission lines or reflection components in the energy spectrum, in analogy with the AMXPs XTE J1807-294 \citep{Falanga05a}, XTE J1751-305 \citep{Miller03}, SWIFT J1756.9-2508 (Sanna et al 2018 submitted) as well as the recently discovered IGR J16597-3704 \citep{Sanna2018}. 
For a source distance of ~8.5~kpc, the observed average flux of $\sim 1.2\times 10^{-11}$ erg s$^{-1}$ cm$^2$ (0.5--10 keV) corresponds to a luminosity of $L=4.3\times10^{35}$ erg s$^{-1}$. 
Assuming accretion-torque equilibrium, we use the latter value to make a rough estimate of the dipolar 
magnetic field B of the NS:
\begin{equation}
\label{eq:spineq}
B=0.2\,\zeta^{-7/4}\left(\frac{P_{\text{spin}}}{2 \,\text{ms}}\right)^{7/6}\left(\frac{M}{1.4M_{\odot}}\right)^{1/3}\left(\frac{\dot{M}}{10^{-11}M_{\odot} \text{yr}^{-1}}\right)^{1/2}10^8 \, \text{G},
\end{equation}
where $\zeta$ is a model-dependent dimensionless factor typically between 0.1 and 1 that describes the relation between 
the magnetospheric radius and the Alfv\'en radius \citep[see e.g.,][]{Ghosh79a,Wang96,Bozzo2009a}, $P_{\text{spin}}$ is NS period in units of ms, $M$ is the NS mass and $\dot{M}$ is the mass accretion rate onto the NS surface. 
Standard NS parameters such as radius $R=10$ km and mass $M=1.4$ M$_{\odot}$, we estimate $0.4\times10^8< B<2.3\times 10^{9}$ G, consistent with the average magnetic field of known AMXPs \citep[see e.g.,][]{Mukherjee2015, Degenaar2017}. 

X-ray pulsations was also detected in the \rxte{} observations of the \igr{} corresponding to its 2004 and 2008 outbursts. We note that orbital Doppler shift corrections were required to unveil the coherent pulsation, implying a relatively poor signal-to-noise ratio. Combining the barycentric spin frequency values observed for each outburst and shown in Tab.~\ref{tab:solution}, we estimated an upper limit (3$\sigma$ c.l.) of the secular spin derivative $-8.3\times 10^{-13}$ Hz/s $< \dot{\nu}<1.1\times 10^{-12}$ Hz/s. Following \citet[][and references therein]{Spitkovsky2006a} we converted the frequency spin-down upper limit into an upper limit on the magnetic field strength of B$<2.8\times 10^{9}$ G (assuming a NS R=10km and an angle $\alpha\simeq 10^{\circ}$ between the magnetic hotspot and the rotational pole), consistent with the estimate reported above.   

Finally, we investigated the orbital period secular evolution of \igr{} by combined the ephemeris of the three observed outbursts of the source, obtaining a more accurate value of the orbital period $P_{orb}$ = 6765.84521(2) s and an orbital period derivative $\dot{P}_{orb}=(-2.5\pm2.3)\times 10^{-12}$s/s. The large uncertainty on the latter finding does not allow to unambiguously determine the secular evolution of the system, however, we notice that within uncertainties the value reported for \igr{} is still compatible with the fast expansion reported for \saxj{} \citep[][]{diSalvo08, Patruno:2017aa, Sanna:2017ab}. Similar results has been reported for the AMXP \swiftj{} (Sanna et al. 2018, submitted). A secular evolution compatible with that observed for the AMXP IGR J00291+5934, that shows evolutionary timescales compatible with conservative mass transfer driven by angular momentum loss via gravitational radiation (GR) \citep{Patruno:2017ab, Sanna2017b}, is however still compatible with our findings. Future outbursts of the source will be of fundamental importance to further constrain the orbital period derivative, hence the secular evolution of the system.

\begin{acknowledgements}
We thank N. Schartel and E. Kuulkers for the possibility to perform the ToO observation in the Director Discretionary Time with \xmm{} and \integral{}, respectively. We also thanks the \xmm{} and \integral{} teams for the rapid scheduling of the ToO observations of \igr{}. A. P. acknowledges funding from the EUs Horizon 2020 Framework Programme for Research and Innovation under the Marie Sk\l{}odowska-Curie Individual Fellowship grant agreement 660657-TMSP-H2020-MSCA-IF-2014.
\end{acknowledgements}

\bibliographystyle{aa} 
\bibliography{biblio.bib}

\end{document}